\documentclass[prd,twocolumn,showpacs,preprintnumbers,superscriptaddress]{revtex4}
\usepackage{epsfig}
\usepackage{amsfonts}
\usepackage{amsmath}
\usepackage{graphicx}

\begin{document}

% Page numbers bottom-center
%\pagestyle{plain}

\title{Fitting CoGeNT Modulation with an Inelastic, Isospin-Violating $Z'$ Model}

\author{Haipeng An}
 \affiliation{Maryland Center for Fundamental Physics and Department of Physics, University of
Maryland, College Park, Maryland 20742, USA }
\author{Fei Gao}
 \affiliation{Department of Physics, Shanghai Jiao Tong University, Shanghai 200240, China}

\date{\today}

\begin{abstract}
We reanalyze the annual modulation data observed by CoGeNT experiment and show that if the annually modulated anomaly detected by CoGeNT is induced by collision between dark matter particle and nucleus, it can be fitted by a $Z'$ model with inelastic dark matter and isospin-violating interaction, and the constraint from XENON100 can be avoided. This $Z'$ model is strongly constrained by collider physics that the upper bound of the mass of $Z'$ is around twice of the mass of dark matter.
\end{abstract}

\maketitle
%\tableofcontents

\section{Introduction}

A lot of astronomical evidence have already shown that more than one fifth of the universe is composed by matter which is not baryonic and does not significantly interact with light, the nature of dark matter (DM), however, remains mysterious. The most compelling version of DM is that it is composed by weakly interacting particles (WIMPs), which offers to understand the relic abundance of DM as natural consequences of the thermal evolution of the universe.

A number of experiments are searching for a signal from WIMPs scattering on nuclei with underground detectors. Among them, CoGeNT collaboration, employing P-type point contact germanium, is able to detect signals with sub-keV nuclear recoils due to their low electronic noise. In early 2010, around 100 events above the expected background were reported by CoGeNT with ionization energy in the range of 0.4 to 1.0 keVee (electron recoil equivalent energy)~\cite{Aalseth:2010vx} with a 56-day exposure in 330 grams germanium inside the fiducial volume. These events can be interpreted as elastic scattering of a 5 $-$ 10 GeV WIMP on nuclei with $\sim10^{-40}$ cm$^{2}$ cross section. Most recently, the CoGeNT collaboration published their new results of 442 live days data~\cite{Aalseth:2011wp}, which confirmed the event excess observed in their previous report and showed a 2.8$\sigma$ indication of an annual modulation of the event rate. However, as noticed in the literature~\cite{Schwetz:2011xm}, the best fit point of the modulation signal is excluded by the unmodulated rate.
On the other hand, the XENON100 collaboration also published their new results from a 4843 kg day exposure~\cite{Aprile:2011ts}, obtaining three events passing all the cuts with an expected background of 1.8$\pm$0.6 events. The large exposure and fiducial volume render XENON100 the most stringent constraint for heavy WIMPs ($\sim$100 GeV). However, the specific energy resolution at low energy region gives XENON100 a strong rejecting power for light WIMPs ($\sim$10 GeV) as well~\cite{Aprile:2010um}, which appears to rule out the light WIMP interpretation of CoGeNT data. However, there are several ways that this conflict can be alleviated. From experimental side, there might be issues with scintillation efficiency of liquid xenon or the quenching factor of the germanium in the region of interest~\cite{Collar:2010nx}. From the theoretical point of view, isospin violating interaction between WIMP and nucleons is proposed~\cite{Feng:2011vu}, where, in particular, if $A_{n}/A_{p}\approx-0.7$, the constraint from xenon-based experiment on germanium based ones would be strongly weakened by a factor of around 20, where $A_{p}$ and $A_{n}$ are the scattering amplitude of WIMP on proton and neutron, respectively.

In summary, it faces two major challenges to interpret the result of CoGeNT as WIMP signals. One is the tension between the unmodulated rate and the modulation from itself, and the other is the conflict from XENON100. The bound on  WIMP-nucleon cross section of XENON100 reduces sharply with the mass of WIMP due to that the nuclear recoil energy is proportional $M_{D}^{2}$ in the case of low mass dark matter, where $M_D$ is the mass of DM, and XENON100 is not very sensitive in low energy region. For example, the upper bound on scattering cross section between WIMP and nucleon weakens by a factor of 20 from $M_{D}=7$ GeV to $M_{D}=6$ GeV.

%On the other hand, if one would like to use elastic scattering of WIMP to explain the observed rate in CoGeNT, the fitted differential scattering spectrum dies off quickly in the range $E_{R}^{e}>0.9$ keVee, where $E_{R}^{e}$ is the effective scintillation energy. However, as is discussed in next section, the annual modulation in the range $E_{R}^{e}>0.9$ keVee is as strong as in the range of 0.5 keVee $<E_{R}^{e}<0.9$ keVee. However, any attempt to use elastic scattering WIMP to fit the tail in $E_{R}^{e}>0.9$ keVee can be excluded by XENON100 since the mass of WIMP must be relatively large. Therefore, a natural candidate to fit the data from 0.9 keVee to 3 keVee is to use inelastic WIMP. However, the inelastic WIMP alone cannot fit the sharp rise of the differential scattering spectrum in the low energy region. Therefore, one need another component of elastic WIMP. However, with the existence of the inelastic WIMP the best fit of the elastic WIMP can be much lighter than before so that the constraint from XENON100 can be avoided. In this work, we propose that if DM in the universe is composed of an elastic WIMP component and an inelastic component as well, the observed signal of CoGeNT can be reconciled with XENON100.

%(Add some references on inelastic dark matter models).

In most prevailing models, the DM candidate has only elastic scattering with nuclei, such as the LSP DM in supersymmetry models~\cite{Goldberg:1983nd} or the KK DM in models with extra dimensions~\cite{Cheng:2002ej}. However, motivated by the annual modulation signal from DAMA/LIBRA experiment~\cite{Bernabei:2008yi}, more attentions have been drawn to the idea of inelastic DM models~\cite{TuckerSmith:2001hy}. After the modulation signal of CoGeNT was announced, people started to think about interpreting the CoGeNT result as signals induced by inelastic WIMP~\cite{Schwetz:2011xm,Fox:2011px}. In this work, we try to fit the CoGeNT modulation data using a leptophobic $Z'$ portal with isospin-violating couplings to quarks in the Standard Model (SM). Furthermore, the DM in this model is a real scatter with a keV energy gap from its excited state, which is generated by a soft breaking of a global U(1) symmetry. Apart from DM direct detection experiments, this model is also strongly constrained from the process of monojet plus missing transverse energy (MET) in Tevatron and LHC.

%If the result of CoGeNT is induced by WIMP, it is difficult to explain it with spin independent elastic WIMP, we even don't know if DM is composed by one component of WIMP or not. Therefore, the usual way, to fit the data from one experiment with some model and then compare it with the exclusion curve of another experiment, is no longer efficient. However, one can directly compare the differential rate from two experiments, or translate the differential rate from one experiment to the other to see if they are consistent or not.

The rest of this work is organized as the following. In Sec.~II, we introduce the basic formulas of DM direct detection for future use. In Sec.~III, we derive a model independent formula to translate the differential rate of WIMP scattering from one experiment to another in the region of low nuclear recoil energy and use it to motivate the idea of inelastic WIMP. In Sec.~IV, we present basic properties of the model and use the model to fit the modulation detected by CoGeNT. In Sec.~V, we discuss the collider constraint on this model. We summarize in Sec.~VI.

\section{Elastic and inelastic WIMPs}

The recoil energy $E_{r}$ of the nucleus hit by WIMP with velocity $v$ in lab frame can be written as
\begin{equation}\label{recoil}
E_{r} = \frac{\mu_{DA}^{2}v^{2}}{M_{A}}\left[ 1-\frac{\delta}{\mu_{DA}v^{2}}-\sqrt{1-\frac{2\delta}{\mu_{DA} v^2}} \cos\theta\right]\ ,
\end{equation}
where $\theta$ is the scattering angle in the center-of-mass frame of the WIMP and nucleus and $\delta$ is the energy gap between WIMP and its excited state. $\mu_{DA}$ is the reduced mass of WIMP and nucleus. Since the mass of WIMP we are interested in in this work is around 10 GeV, the values of $\mu_{DA}$ in Germanium and Xenon are similar to each other. Therefore, from Eq.~(\ref{recoil}), one can see that the dependence on target nucleus relies on $m_{A}$, and the dependence is quite simple so that one can translate the differential energy spectrum from one detector to the other.

The differential scattering rate of WIMP with mass $M_{D}$ and velocity distribution $f({\mathbf v})$ can be written as
\begin{equation}\label{drder}
\frac{d R}{d E_{r}} = \frac{\rho_{D}}{M_{D}}\frac{1}{M_{A}}\int_{| {\mathbf v} |>v_{\rm min}}d^{3}v\frac{d\sigma_{A}}{d E_{r}} |{\mathbf v}| f({\mathbf v})\ ,
\end{equation}
where $M_{A}$ and $E_{r}$ are the mass and the recoil energy of the target nucleus, and $v_{\rm min}$ is the minimal velocity for the WIMP to generate certain nuclear recoil energy $E_{r}$, which can be written as
\begin{equation}\label{vmin}
v_{\rm min} = \frac{1}{\sqrt{2M_{A}E_{r}}}\left(\frac{M_{A} E_{r}}{\mu_{DA}}+\delta\right)\ ,
\end{equation}
where $\delta$ is the energy gap between DM and its lowest excited state, $\delta = 0$ in the case of elastic scattering.

The differential cross section of WIMP and nucleus in Eq.~(\ref{drder}) can be written as
\begin{equation}\label{dsder}
\frac{d\sigma_{A}}{d E_{r}} = \frac{M_{A}\bar\sigma_{N}}{2\mu_{DN} v^{2}} A^{2}_{\rm eff} F^{2}({\mathbf q}^{2})\ ,
\end{equation}
where $\mu_{DN}$ is the reduced mass of WIMP and nucleon, $\bar\sigma_{N}\equiv(\sigma_{n}+\sigma_{p})/2$ is the average scattering cross section between WIMP and nucleon and
\begin{equation}
A^{2}_{\rm eff} = \sum_{i\in {\rm isotopes}} 2r_{i}[Z\cos\theta_{N} + (A_{i}-Z)\sin\theta_{N}]^{2}\ ,
\end{equation}
where $r_{i}$ are relative abundances of isotopes, $\tan\theta_{N}\equiv A_{n}/A_{p}$. In Eq.~(\ref{dsder}), $F({\mathbf q}^{2})$ is the form factor of nucleus, and in this work, we assume $F_{n} = F_{p} = F$ and use form factor proposed by Helm~\cite{Helm:1956zz} that
\begin{equation}
F(q r_{n}) = 3 \frac{j_{1}(q r_{n})}{qr_{n}}\times e^{-(qs)^{2}/2} \ ,
\end{equation}
where $j_{1}$ is the first order spherical Bessel function and $r_{n}$ is an effective nuclear radius can be written as
\begin{equation}
r_{n}^{2} = c^{2} + \frac{7}{3}\pi^{2}a^{2} - 5s^{2}\ ,
\end{equation}
where $c\simeq(1.23 A^{1/3}-0.60) {~\rm fm}$ and $s=0.9$ fm.

The differential scattering rate in Eq.~(\ref{drder}) depends strongly on the velocity distribution of dark matter halo. In this work we assume a standard halo model with the isotropic Maxwellian velocity distribution:
\begin{equation}
f(\mathbf{v}) = \frac{n_{0}}{k}\left\{ \begin{array}{cc}  \exp[-(\mathbf{v}+\mathbf{v}_{\rm E})^{2}/v_{0}^{2}]\;,~&~ |\mathbf{v}+\mathbf{v}_{\rm E}|<v_{\rm esc} \\ 0\;, ~&~ |\mathbf{v}+\mathbf{v}_{E}|\geq v_{\rm esc}\end{array}\right.\ ,
\end{equation}
where $\mathbf{v}_{\rm E}$ is the velocity of the earth, $v_{\rm esc}$ is the local Galactic escape velocity and $k$ is the normalization factor which is
\begin{equation}
k = (\pi v_{0}^{2})^{3/2}\left[ {\rm erf}\left(\frac{v_{\rm esc}}{v_{0}}\right)-\frac{2}{\pi^{1/2}}\frac{v_{\rm esc}}{v_{0}}e^{-v^{2}_{\rm esc}/v_{0}^{2}} \right]\ ,
\end{equation}
where $\rm erf$ is the error function and $v_{\rm esc}$ is the escape velocity of DM in local galaxy which is about 544 km/s~\cite{Lewin:1995rx}.

The velocity distribution of DM is formed due to gravitational effect. If the collision between DM and other regular matter in the universe is dominated by inelastic scattering, the kinetic energy of the system gets smaller during the collision. Therefore, if there are two components of DM in the same halo, one is elastic and the other is inelastic, one may expect that the average velocity of the inelastic DM is smaller than the elastic one. Therefore, in this work, we choose a relatively small $v_0$ which is 200 km/s.

$f(\mathbf{v}_{\rm E})$ is modulated annually due to the rotation of the earth around the sun, which then induces the annual modulation of the scattering rate between WIMP and nucleus. The velocity of the earth can be parameterized as the following~\cite{Lewin:1995rx}:
\begin{equation}
v_{\rm E} = v_{\rm E}^{(0)} + v_{\rm E}^{(1)}\sin[2\pi (t-t_{0}) / 1{~\rm year}]\ ,
\end{equation}
where $v_{\rm E}^{(0)}=232$ km/s and $v_{\rm E}^{(1)}=15$ km/s.

Due to the sensitivity limit of the detector and the nonlinear velocity distribution, the pattern of annual modulation of the scattering rate induced by the velocity of the earth can be seen as a deformed sine wave. To quantify it, we define the modulation as
\begin{equation}\label{modulation}
{\cal M} = \frac{A_{1}}{A_{0}}\ ,
\end{equation}
where $A_{0}$ and $A_{1}$ stand for the amplitudes of zeroth and first Fourier modes of the scattering rate, respectively. The exact definitions of $A_0$ and $A_1$ and the uncertainty of ${\cal M}$ are studied in Appendix A.

\section{CoGeNT data with constraint from XENON100}

\subsection{Relations between CoGeNT and XENON100}

\begin{figure}[h]
\includegraphics[width=3in]{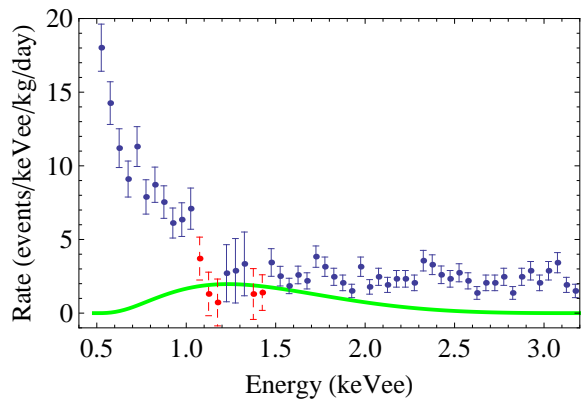}
\caption{Total differential rate observed by CoGeNT with the prediction from the inelastic WIMP model with $m_\chi = 13.3$ GeV, $\Delta m_\chi = 33$ keV, $\bar\sigma_N\times {(A_{\rm eff}^{\rm Ge})}^2 = 6\times10^{-38}$ cm$^2$.}
\label{fig:CoGeNT_bg_spectrum}
\end{figure}

The differential scattering rate measured by CoGeNT with 435 days data of 330 gram fiducial volume is shown in Fig.~\ref{fig:CoGeNT_bg_spectrum}, where the dots represent the spectrum with the L-shell electron capture (EC) background subtracted using the method in Ref.~\cite{Fox:2011px}. The errorbars include only the statistic uncertainties. After subtracting the L-shell EC background, we found that the differential rate shown in the bins labeled by red dots in Fig.~\ref{fig:CoGeNT_bg_spectrum} are obviously smaller than expected, which may be caused by some unknown systematic errors of the subtraction. Therefore, the information of these five bins is not used for fitting the unknown background.

%\FIGURE[th]{
%\includegraphics[scale=0.6]{CoGeNT_bg_spectrum.pdf}
%\caption{ Dominant parton level diagrams for $p\bar p\rightarrow$ monojet + MET.  (b) is also the dominant parton level process for $pp\rightarrow$ monojet + MET in LHC.
%\label{fig:CoGeNT_bg_spectrum}}}

As was noticed in Refs.~\cite{Fox:2010bz,Fox:2011px}, one can do a halo independent comparison between different direct detection experiments. From Eqs.~(\ref{drder}) and (\ref{dsder}), the differential scattering rate can be written as
\begin{equation}\label{drderA}
\frac{dR}{d E_{r}} = \frac{\rho_{D}}{M_{D}} \int _{|\mathbf{v}|>v_{\rm min}} d^{3}\mathbf{v} \frac{\bar\sigma_{N}}{2\mu_{DA}v}A^{2}_{\rm eff} F^{2}(|\mathbf{q}|) f(\mathbf{v})\ .
\end{equation}
In the case of small nuclear recoil where we are interested in, since the target nucleus can be seen as rigidity, we have $F(|\mathbf{q}|)\approx 1$. Furthermore, if these nuclear recoils are generated by scattering of WIMPs, the mass of these WIMPs must be around 10 GeV, which is much smaller than the masses of germanium and xenon. Therefore, the reduced mass $\mu_{DA}$ in Eq.~(\ref{drderA}) can be replaced by $M_{D}$. With these simplifications, Eq.~(\ref{drderA}) can be written as
\begin{equation}
\frac{dR}{dE_{r}} \approx \frac{\rho_{D}\bar\sigma_{N}A^{2}_{\rm eff}}{2m_{D}^{2}}\int_{|\mathbf{v}|>v_{\rm min}(E_{r})} d^{3}\mathbf{v} \frac{f(\mathbf{v})}{v}\ .
\end{equation}
From Eq.~(\ref{vmin}), with the approximation $\mu_{DA}\approx M_{A}$, one can see that $v_{\rm min}$ depends only on the product $E_{r} M_{A}$. Therefore, we can get
\begin{equation}\label{translator}
\frac{d R}{d(M_{A}E_{r})} \approx \frac{\rho_{D}\bar\sigma_{N}}{2m_{D}^{2}}\frac{A^{2}_{\rm eff}}{M_{A}}\int_{|\mathbf{v}|>v_{\rm min}(M_{A}E_{r})} d^{3}\mathbf{v} \frac{f(\mathbf{v})}{v}\ ,
\end{equation}
where one can see that the only dependence of $d R/d(M_{A}E_{r})$ on the target nucleus is in the factor $A^{2}_{\rm eff}/M_{A}$. Therefore, Eq.~(\ref{translator}) can be used to translate differential scattering spectrum between different direct detection experiments. As a result, the relation between differential scattering rates of CoGeNT and XENON100 can be written as
\begin{equation}
\left.\frac{M_{A}}{A^{2}_{\rm eff}}\frac{d R}{d(M_{A}E_{r})}\right|_{\rm Ge} \approx \left.\frac{M_{A}}{A^{2}_{\rm eff}}\frac{d R}{d(M_{A}E_{r})}\right|_{\rm Xe}\ .
\end{equation}
Therefore, one can get that
\begin{equation}
\frac{d R^{\rm Xe}}{d E_{r}^{\rm Xe}} = \left(\frac{A_{\rm eff}^{\rm Xe}}{A_{\rm eff}^{\rm Ge}}\right)^{2} \frac{d R^{\rm Ge}}{d E_{r}^{\rm Ge}}\ ,
\end{equation}
where
\begin{equation}
E_{r}^{\rm Ge} = \frac{M_{\rm Xe}}{M_{\rm Ge}} E_{r}^{\rm Xe}\ .
\end{equation}

\subsection{CoGeNT annual modulation}

\begin{figure}[!h]
\includegraphics[width=3in]{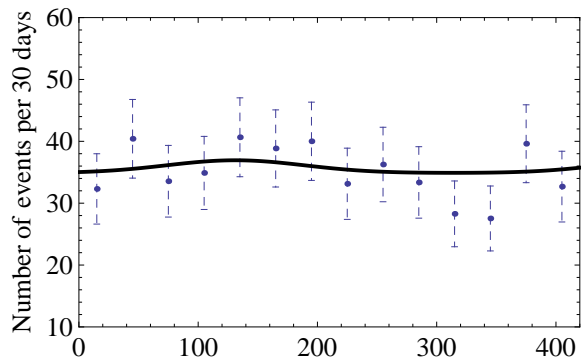} \\
(a)\\
\includegraphics[width=3in]{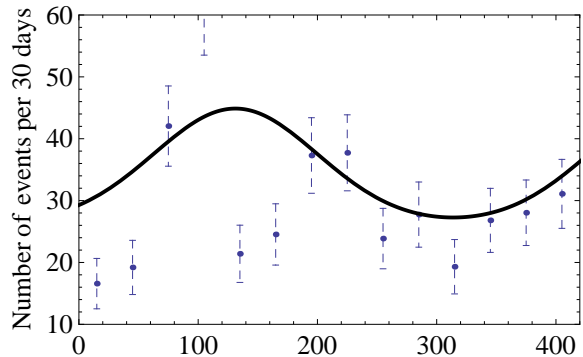} \\
(b)\\
\includegraphics[width=3in]{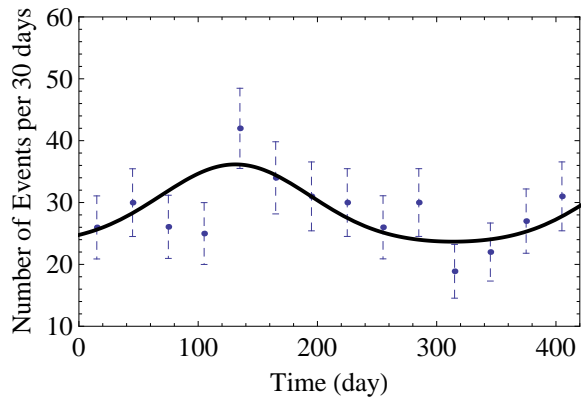} \\
(c)\\
\caption{Fitting the annual modulation signal detected by CoGeNT. The data points represent the observed single after subtracting the L-shell EC background. The black curves show the prediction of the $Z'$ model with $m_\chi = 13.3$ GeV, $\Delta m_\chi = 33$ keV, $\bar\sigma_N\times {(A_{\rm eff}^{\rm Ge})}^2 = 6\times10^{-38}$ cm$^2$. (a) is for ionization energy from 0.5 to 0.8 keVee, (b) from 0.8 to 1.5 keVee and (c) from 1.6 to 2.8 keVee.}
\label{fig:modulation_new_1}
\end{figure}

The annual modulation spectra are shown in Fig.~\ref{fig:modulation_new_1} for recoil energy in three different energy regions. One can see that the amplitude of modulation is still around 15\% in the region of 1.6 keVee to 2.8 keVee. Therefore, if we assume the only source of annual modulation observed by CoGeNT is the scattering of WIMPs, part of the measured spectrum in the region of 1.6 keVee to 2.8 keVee shown in Fig.~\ref{fig:CoGeNT_bg_spectrum} should include WIMP signal. The quenching factor in this region is around 0.25 and the cut efficiency is about 0.9. From Fig.~\ref{fig:CoGeNT_bg_spectrum} we can get that the measured spectrum in this region is also around a constant which is about 3 $/{\rm keVee{\rm /kg/day}}$. Therefore, using Eq.~(\ref{translator}), we can get that the corresponding range for xenon detector is about 3.7 keVnr to 6.0 keVnr with differential scattering spectrum
\begin{equation}
\left.\frac{d R}{d E_{r}}\right|_{\rm Xe} \approx 0.8\times \left(\frac{A_{\rm eff}^{\rm Xe}}{A_{\rm eff}^{\rm Ge}}\right)^{2} {\rm /keVnr/kg/day}\ .
\end{equation}
Therefore, with the Poisson smearing effect~\cite{Aprile:2010um}, the distribution of S1 signal in XENON100 detector can be written as
\begin{equation}\label{corresponding}
N_{S1}(n) = \eta(n)\int_{E_{r}^{\rm min}}^{E_{r}^{\rm max}} d E_{r} {\rm Poisson}(n,\mu_{S1}(E_{r})) \frac{d R}{d E_{r}}\ ,
\end{equation}
where $\eta$ is the acceptance of S1 signal in XENON100, $\mu_{S1} = E_{\rm nr} L_{\rm y} L_{\rm eff} S_{\rm nr}/S_{\rm ee}$ is the expectation value of the number of photoelectrons induced by nuclear recoil $E_{r}$. The values of the quantum yielding efficiency $L_{\rm y}$, the scintillation efficiency $L_{\rm eff}$ and the electric scintillation quenching factors $S_{\rm ee}$ and $S_{\rm nr}$ can be got from Ref.~\cite{Aprile:2011ts}.

In XENON100, the energy window of WIMP search region is chosen between 4-30 photoelectrons. From Eq.~(\ref{corresponding}), we can get that the spectrum in the region of 1.5 keVee to 2.5 keVee observed by CoGeNT is corresponding to
\begin{equation}
N \approx 290\times ~{\rm events}~ \times \left(\frac{A_{\rm eff}^{\rm Xe}}{A_{\rm eff}^{\rm Ge}}\right)^{2}\ .
\end{equation}
If the interaction between WIMP and nucleons is isospin-violating, $({A_{\rm eff}^{\rm Xe}}/{A_{\rm eff}^{\rm Ge}})^{2}$ can be as small as $1/20$. Therefore, we can get that, if CoGeNT spectrum in the region of 1.5 keVee to 2.5 keVee were all induced by WIMP, XENON100 should have already observed around 10 events. However, XENON100 reported only 3 events passed all cuts and only one of them lies in the low energy region with six photoelectrons observed. Therefore, we can get the conclusion that WIMP signals amounts at most to 10\% of scatterings in the region of 1.6 keVee to 2.8 keVee.

However, as shown in Fig.~\ref{fig:modulation_new_1}(c), the modulation of the data is around 15\%, which means either the background itself has an annual modulation with the same phase as the one generated by WIMP, or the modulation of WIMP is around 100\% with the background having no modulation at all. In this work we pursue the second possibility that we assume the background has no dependence on time and all the annual modulation of the CoGeNT data is generated from WIMP. To generate 100\% annual modulation, we need to resort to a DM model in which the interaction between WIMP and nucleus is inelastic and isospin-violating as well.

Since only a small fraction of the differential rate shown in Fig.~\ref{fig:CoGeNT_bg_spectrum} can be generated by WIMP. We only use our model to fit the modulation and use the total rate as a constraint.

%\FIGURE[th]{
%\includegraphics[scale=1.5]{modulation_new_1.pdf}
%\caption{ Dominant parton level diagrams for $p\bar p\rightarrow$ monojet + MET.  (b) is also the dominant parton level process for $pp\rightarrow$ monojet + MET in LHC.
%\label{fig:modulation_new_1}}}

%\FIGURE[th]{
%\includegraphics[scale=1.5]{modulation_new_2.pdf}
%\caption{ Dominant parton level diagrams for $p\bar p\rightarrow$ monojet + MET.  (b) is also the dominant parton level process for $pp\rightarrow$ monojet + MET in LHC.
%\label{fig:modulation_new_2}}}

%\begin{figure}[!h]
%\includegraphics[width=3in]{modulation_new_2.eps}
%\caption{}
%\label{fig:modulation_new_2}
%\end{figure}

\section{A simple model for isospin-violating inelastic WIMP}

Theoretically, it is not difficult to get an inelastic WIMP with isospin-violating interaction. The isospin-violating $Z'$ model has already been proposed in the literature~\cite{Frandsen:2011cg}. In this work, we extend this model to inelastic WIMP. The Lagrangian can be written as the following,
\begin{eqnarray}\label{lagrangian}
{\cal L} &=& Z'_\mu \left[\left( g_u \bar U_R \gamma^\mu U_R + g_d \bar D_R \gamma^\mu D_R \right)\right. \nonumber \\
&& + \left.g_D  i(\chi^\dagger \partial^\mu \chi - \partial^\mu\chi^\dagger \chi) \right]\ ,
\end{eqnarray}
where $g_u$ and $g_d$ are couplings between $Z'$ to up-type and down-type quarks, respectively. $\chi$ is the candidate of DM stabilized by a global $U(1)$ symmetry with mass $m_\chi$. Now, consider the case that this $U(1)$ symmetry is softly broken by the following mass terms
\begin{equation}
 - \frac{1}{2}\delta^2 [\chi^2 + {\chi^\dagger}^2] \ .
\end{equation}
where $\delta^2$ is assumed to be real. It is technically natural to assume $\delta^2 \ll m_\chi^2$, since it is a parameter which softly breaks a $U(1)$ global symmetry. Then, a mass gap $\Delta m_\chi = \delta^2/m_\chi$ is generated between the real and imaginary parts of $\chi$. Therefore, for a 10 GeV dark matter to get a mass gap around 30 keV, the symmetry breaking parameter $\delta$ is around 20 MeV. In this work, we use $\chi_1$ and $\chi_2$ to label DM and its excited state, respectively.

In the Lagrangian shown in Eq.~(\ref{lagrangian}), to avoid flavor changing neutral current, the couplings $g_u$ and $g_d$ should be universal for all the three families of quarks. Then, it is easy to see that
\begin{equation}
\frac{A_n}{A_p} = \frac{g_u + 2g_d}{2g_u + g_d} \ ,
\end{equation}
where $A_n$ and $A_p$ are scattering matrix elements of WIMP to neutron and proton, respectively. One can see that if $g_u/g_d \approx -1.125$, $A_n/A_p$ approaches to $-0.7$, where the sensitivity of xenon is significantly reduced compared to germanium. The cross section between WIMP and nucleons can be written as
\begin{equation}
\sigma_{p(n)}=\frac{(2g_{u(d)}+g_{d(u)})^2 g_D^2 m_{p(n)}^2m_\chi^2}{4\pi M_{Z'}^4 (m_{p(n)}+m_\chi)^2}\ .
\end{equation}

It is easy to see that this $U(1)$ symmetry is not anomalous free, however, one can always introduce heavy spectating fermions into the model to free the anomaly, and the upper bound of mass of the spectators is around $64\pi^2 M_{Z'} / g_{Z'}^3$~\cite{Preskill:1990fr}, which can be far away from the reach of detectors.

%\subsection{Low energy region}

%From Fig.~\ref{drder}, we can see that the differential scattering rate of CoGeNT in low energy region sharply goes up. However, the differential rate induced by inelastic WIMP goes down at low energy region. Furthermore, as the recoil energy becomes smaller, the modulation also becomes smaller. Therefore, it is difficult for inelastic dark matter to generate large modulation at low energy region. However, in the energy region of $0.5$~keVee to $0.9$~keVee, from Fig.~\ref{fig:modulation_new_1} one can see that the modulation is around 20\%. Therefore, to fit the modulation data, we introduce another component of WIMP and the scattering with nuclei is assumed to be elastic.

%\FIGURE[th]{
%\includegraphics[scale=1.5]{rate.pdf}
%\caption{ Dominant parton level diagrams for $p\bar p\rightarrow$ monojet + MET.  (b) is also the dominant parton level process for $pp\rightarrow$ monojet + MET in LHC.
%\label{fig:rate}}}

%\begin{figure}[!h]
%\includegraphics[width=3in]{rate.eps}
%\caption{}
%\label{fig:rate}
%\end{figure}

%\section{Fit the modulation data}

From the discussion of Sec.~IIB, we have already seen that if one wants to interpret the annual modulation of CoGeNT as signal from WIMP, a large amount of events should be unknown background other than the known L-shell EC background. Therefore, we do not attempt to use the model to fit the differential rate in Fig.~\ref{fig:CoGeNT_bg_spectrum}. Our strategy to calculate the annual modulation is the following. We first get an interpolating function of differential rate with the L-shell EC background subtracted. The model predicts a differential rate for CoGeNT, taking the green curve in Fig.~\ref{fig:CoGeNT_bg_spectrum} as an example. Then, we subtract the predicted differential rate from the interpolating function to get the unknown background. Assuming the unknown background has no annual modulation we can calculate the modulation predicted by the model.

To fit the observed modulation of CoGeNT, we need the modulation from WIMP to be around 100\%. In reality, the parameters we use are $m_\chi = 13$ GeV, $\Delta m_\chi = 33$ keV, $\bar\sigma_N\times {(A_{\rm eff}^{\rm Ge})}^2 = 6\times10^{-38}$ cm$^2$. The predicted differential rate is shown by the green curve in Fig.~\ref{fig:CoGeNT_bg_spectrum}, where one can see that it does not exceed the observed one.

\begin{figure}[!h]
\includegraphics[width=3in]{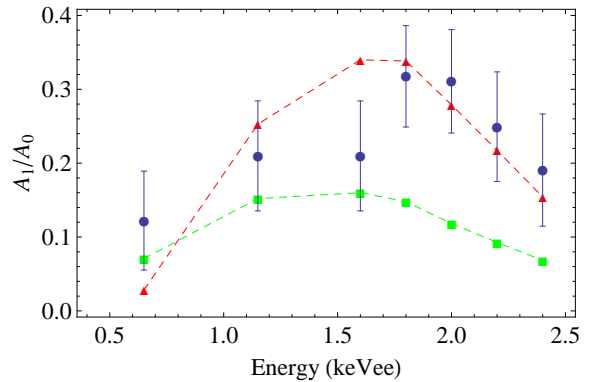}
\caption{Annual modulation as a function of ionization energy. The blue dots show the modulation extracted from CoGeNT experiment after subtracting the L-shell EC background, while the red curve shows the corresponding theoretical predictions from the model discussed in Sec.~IV with the parameters chosen as the same as in the caption of Fig.~\ref{fig:modulation_new_1}. The energy region of the first plot is from 0.5 to 0.8 keVee, the second from 0.8 to 1.5 keVee, the third from 1.0 to 2.2 keVee, the fourth from 1.2 to 2.4 keVee, the fifth from 1.4 to 2.6 keVee, the sixth from 1.6 to 2.8 keVee and the seventh from 1.8 to 3.0 keVee. The dashed green curve shows the corresponding predictions using the parameters from Ref.~\cite{Schwetz:2011xm}.}
\label{fig:modulation_energy}
\end{figure}

The predicted curves of modulation for different energy regions of 0.5 $-$ 0.8 keVee, 0.8 $-$ 1.5 keVee, and 1.6 $-$ 2.8 keVee are shown in Fig.~\ref{fig:modulation_new_1} together with the binned CoGeNT data. The change of the modulation with the energy is shown in Fig.~\ref{fig:modulation_energy}, where the predicted data are represented with red triangles while the experimental data are shown with blue round dots. The error bars shown in the plot include only the statistic uncertainties, which is discussed in Appendix A. The energy ranges for the seven points shown in the plots are 0.5 $-$ 0.8 keVee, 0.8 $-$ 1.5 keVee, 1.0 $-$ 2.2 keVee, 1.2 $-$ 2.4 keVee, 1.4 $-$ 2.6 keVee, 1.6 $-$ 2.8 keVee, and 1.8 $-$ 3.0 keVee, respectively. Since the energy regions overlap with each other, the last six points are not independent so that one cannot do a $\chi^2$ analysis using the plot in Fig.~\ref{fig:modulation_energy}. The reason for choosing overlapped regions is that, the events are not
sufficient enough. The energy bins we are choosing are to keep the average number of events per month to be not smaller than 30, so that from Eq.~(\ref{uncertainty}) the statistic uncertainty is smaller than 7.5\%.

One can see that the theoretical prediction of the annual modulation follows exactly the trend of the experimental data. The modulation is small at the first point. The reason is that the inelastic scattering rate diminishes at very small recoil energy as shown in Fig.~\ref{fig:CoGeNT_bg_spectrum}. In the energy region of 1.0 $-$ 1.5 keVee, the predicted rate reaches its maximum. As a result the modulations also become large. In the energy region of around 1.5 keVee or larger, although the differential rate gets suppressed. The modulation signal is still strong due to that higher velocity is needed to generate large nuclear recoil. The dashed green curve stands for the corresponding prediction with parameters used in Ref.~\cite{Schwetz:2011xm}, where we can see that in high energy region it is lower than the experimental value.

With these parameters, the number of events that XENON100 should have seen in the region of 4 to 12 photoeletrons is about
\begin{equation}
50\times\left(\frac{A^{\rm Xe}_{\rm eff}}{A_{\rm eff}^{\rm Ge}}\right)^2 \ .
\end{equation}
In this energy region, only one event passed all the cuts with the expected background to be 0.6. Using the Feldman-Cousins method~\cite{Feldman:1997qc}, to be consistent with the XENON100 result up to 90\% C.L., the predicted number of events should be smaller than 3.6. Therefore, XENON100 constraint can be avoided by choosing $(A^{\rm Xe}_{\rm eff}/A^{\rm Ge}_{\rm eff})^2 < 0.072$, which requires
\begin{equation}\label{AnAp}
-0.706 < A_n/A_p < -0.674 \ .
\end{equation}

CDMS also uses germanium as target materials, which may render a strong constraint on CoGeNT result. Conflicts from the scattering rate can no longer be alleviated using the idea of isospin-violating interactions. However, there might be issues related to the calibration of CDMS energy scale of the CDMS detector~\cite{Collar:2011kf}, and the different temperatures in CDMS and CoGeNT may also cause a discrepancy between their detecting ability. Therefore, in this work, we are not going to take the constraint from CDMS into consideration.

\section{Collider Constraint}

One can use the process
\begin{equation}\label{ppbar}
p\bar p \rightarrow {\rm monojet+}\chi_1\chi_2\ ,
\end{equation}
in Tevatron to constrain this model. $\chi_2$ may or may not decay inside the collider. However, since the mass gap between $\chi_1$ and $\chi_2$ is $\sim$ 30 keV, $\chi_2$ can decay to $\chi_1$ by emitting only photons or neutrinos or some unknown light particles in a hidden sector. Both neutrinos and hidden light particles cannot be seen by the detector and the photons are also too soft to be detected. Therefore, the signal of this process is a single jet plus missing transverse energy (MET).

\begin{figure}[!h]
\includegraphics[width=3in]{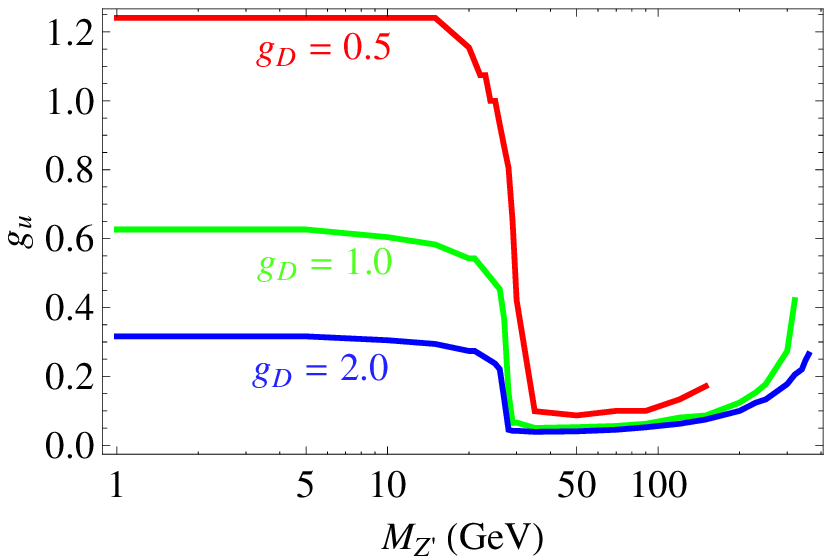} \\
(a)\\
\includegraphics[width=3in]{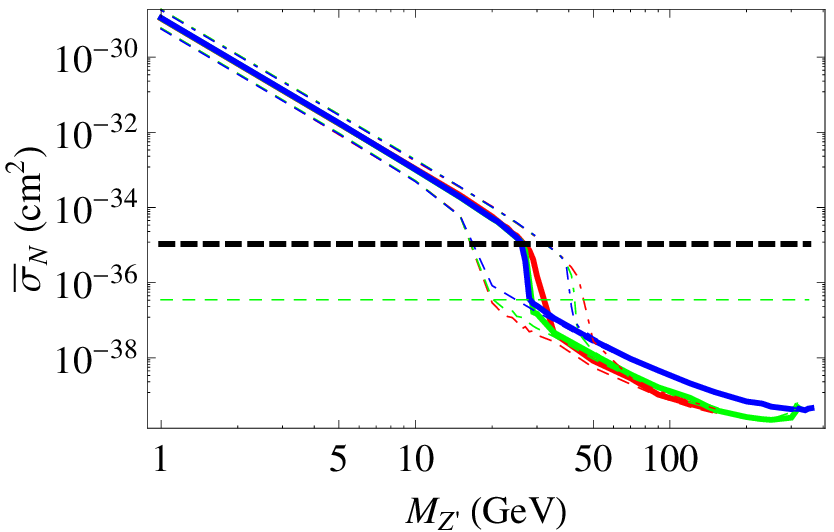} \\
(b)\\
\caption{Collider constraints on isospin violating $Z'$ model. (a) shows the constraint on $g_u$ with $g_D$ fix to be 0.5 (red), 1.0 (green) and 2.0 (blue). (b) shows the corresponding constraint on $\bar\sigma_N$. The thin dashed curves and dotted dashed curves are for $m_\chi = 7.5$ and 20 GeV, respectively. The thick dashed black horizontal line is the value used to fit CoGeNT modulation in this work. The green thin dashed horizontal line shows $\sigma_N$ used in Ref.~\cite{Schwetz:2011xm}.}
\label{fig:collider}
\end{figure}

The CDF group used data with a luminosity of 1 fb$^{-1}$ to study this process~\cite{Aaltonen:2008hh}. With the cut that both the transverse momentum ($P_T$) of the leading jet and the MET should be larger than 80 GeV, $P_T$ of the second hardest jet should be smaller than 30 GeV and vetoing any third jet with $P_T$ larger than 20 GeV, 8449 events were found with the expected SM background to be $8663\pm332$. To set a $2\sigma$ limit on new physics, we require that the cross section for the process in Eq.~(\ref{ppbar}) should be smaller than 0.664 pb. In this work, we work on the parton level process. Pioneered works have already been done in the literature~\cite{collider}. A detailed study of $Z'$ model will be published elsewhere~\cite{anjiwang}. Here, we only present the result of the isospin-violation case. To satisfy the condition in Eq.~(\ref{AnAp}), we can get $-1.139<g_u/g_d<-1.123$. Therefore, in the study of collider constraint, we fix this ratio to be  $-1.125$ which gives the strongest isospin-violating effect of xenon and germanium. The upper bounds for $g_u$ with different values of $g_D$ are shown in Fig.~\ref{fig:collider} (a), with the corresponding upper bounds on $\bar\sigma_N$ shown in Fig.~\ref{fig:collider} (b). The sharp cliffs at around $2m_\chi$in (a) are caused by the transition of $Z'$ from off-shell to on-shell. The black dashed curve in (b) shows the $\bar \sigma_N$ required to generate the modulation signal. One can see that to satisfy the constraints from direct detection and colliders, the upper bound on $M_{Z'}$ is around twice of $m_\chi$. This upper bound is quite common in the case of isospin-violating, inelastic models. The reason is that, to get a relative enhancement of germanium over xenon, ${A^{\rm Ge}_{\rm eff}}^2$ is as small as $5\times10^{-3}$. Furthermore, in the case of inelastic scattering, only a very small fraction of DM particles has enough kinetic energy to interact with the target nucleus. This fraction is around 0.01 in the case with parameters in this work and around 0.1 in Ref.~\cite{Schwetz:2011xm}. Therefore, the constraint from collider physics is enhanced by a factor of $10^{4}\sim10^{5}$. Therefore, in this case, it is a generic feature that $Z'$ should be produced in the collider so that the cross section for the process in Eq.~(\ref{ppbar}) is suppressed by the phase space of three body final state.

\section{Summary and discussions}

In this work, we use a simple $Z'$ model with inelastic WIMP to fit the modulation data observed by CoGeNT. We found that with a relatively small $v_0$, a 13.3 GeV inelastic WIMP with 33 keV mass gap from its excited state can be used to fit the modulation data and avoid the constraint from XENON100. We considered the constraint from Tevatron on this model, and find that $M_{Z'}$ should be smaller than $2 m_\chi$.

As noticed in Ref.~\cite{Finkbeiner:2009mi}, if $\chi_2$ can only decay into $\chi_1$ through kinetic mixing between $Z'$ and $B$, where $B$ is the gauge boson corresponding to $U(1)_Y$ gauge group in SM, the life time of $\chi_2$ will be longer than the age of the universe and the ``down-scattering'' of $\chi_2$ on the nuclei will give the dominant contribution to nuclear recoil. As a result, the nuclear recoil tends to be larger compared to the ``up-scattering'' case so that the constraint from XENON 100 would be more stringent. Furthermore, in the ``down-scattering'' case, since the internal energy released from $\chi_2$ can be converted into the kinetic energy of the recoiled nucleus, the dependence of the nuclear recoil on the velocity of DM is not as strong as in the case of ``up-scattering''. Therefore, it is not easy for the ``down-scattering'' collision to generate enough annual modulation to fit the CoGeNT data. As a result, one needs to invoke other mechanisms to make the decay of $\chi_2$ more rapid.

%\FIGURE[th]{
%\includegraphics[scale=1.5]{modulation_energy.pdf}
%\caption{ Dominant parton level diagrams for $p\bar p\rightarrow$ monojet + MET.  (b) is also the dominant parton level process for $pp\rightarrow$ monojet + MET in LHC.
%\label{fig:monojet-diagram}}}

\acknowledgments

This work was partially supported by the U.~S.~Department of Energy via grant DE-FG02-93ER-40762. We appreciate J.~Collar for providing us the original data of CoGeNT experiment. We also would like to thank L.-T.~Wang, X.-D.~Ji, K.~Ni and H.-B.~Yu for useful discussions.

\appendix

\section{Statistic uncertainty of modulation}

The Fourier transformation of signal can be written as
\begin{eqnarray}\label{fourier}
A_0 &=& \frac{1}{T} \int_0^T F(t) dt\ ;\nonumber \\
A^s_1 &=& \frac{2}{T} \int_0^T F(t) \sin (2\pi t/T) dt\ ; \nonumber \\
A^c_1 &=& \frac{2}{T} \int_0^T F(t) \cos (2\pi t/T) dt\ ; \nonumber \\
A_1 &=& \sqrt{{A^s_1}^2+{A^c_1}^2}\ ,
\end{eqnarray}
where $F$ is the observed differential rate and $T$ is one year. In the analysis, to get a good statistics, each point in Fig.~\ref{fig:modulation_new_1} represents a collection of the data in a month. Therefore, the Fourier analysis in Eq.~(\ref{fourier}) should be discretized that
\begin{eqnarray}\label{fourier_D}
A_0 &=& \frac{1}{N}\sum_{i=1}^N F_i \ ;\nonumber\\
A^s_1 &=& \frac{2}{N} \sum_{i=1}^N F_i \sin (2\pi iT/N) \ ; \nonumber \\
A^c_1 &=& \frac{2}{N} \sum_{i=1}^N F_i \cos (2\pi iT/N) \ ; \nonumber \\
A_1 &=& \sqrt{{A^s_1}^2+{A^c_1}^2}\ ,
\end{eqnarray}
where $N$ is the number of partition in one year.

If all the signals observed are induced by noise, one may still observe an annual modulation using the Eq.~(\ref{modulation}). In this case, we can assume $F_i$ in Eq.~(\ref{fourier_D}) obeys a Poisson distribution with expectation value $\bar F$. Then, it is easy to get that
\begin{eqnarray}
\langle A^s_1\rangle &=& \langle A^c_1 \rangle = 0\ ,\nonumber\\
\langle {A^s_1}^2 \rangle &=& \langle {A^c_1}^2 \rangle = 2\bar F / N \ .
\end{eqnarray}
Therefore, the statistical uncertainty of the modulation can be written as
\begin{equation}\label{uncertainty}
\delta {\cal M} = \sqrt{\frac{2}{N\bar F}}\ .
\end{equation}

%%%%%%%%%%%%%%%%%%%%%%%%%%%%%%%%%%%%%%%%%%%%%%%%%

\end{document}